\documentclass[journal]{IEEEtran}

\usepackage{caption}
\usepackage{subcaption}
\ifCLASSINFOpdf
  \usepackage[pdftex]{graphicx}
  \graphicspath{{./graph/}}
  \DeclareGraphicsExtensions{.pdf,.jpeg,.png}
\else
  \usepackage[dvips]{graphicx}
  \graphicspath{{./graph/}}
  \DeclareGraphicsExtensions{.eps}
\fi

\ifCLASSINFOpdf
\else
\fi

\usepackage[cmex10]{amsmath,mathtools}
\usepackage{mdwmath}
\usepackage{mdwtab}
\usepackage{amssymb}
\usepackage{tabularx}
\usepackage{algorithmic}
\usepackage{algorithm}
\usepackage{float}
\usepackage{booktabs}

\usepackage{epstopdf}
\usepackage{color}
\usepackage{diagbox}
\usepackage{cite}

\ifodd 1

\else
\fi

\begin{document}

\title{Towards Ubiquitous Semantic Metaverse: Challenges, Approaches, and Opportunities}

\author{Kai~Li,~\IEEEmembership{Senior Member,~IEEE,}
        Billy~Pik~Lik~Lau,~\IEEEmembership{Member,~IEEE,}
        Xin~Yuan,~\IEEEmembership{Member,~IEEE,}
        Wei~Ni,~\IEEEmembership{Senior Member,~IEEE,}
        Mohsen~Guizani,~\IEEEmembership{Fellow,~IEEE,}
        and~Chau~Yuen,~\IEEEmembership{Fellow,~IEEE}
\thanks{K.~Li is with Real-Time and Embedded Computing Systems Research Centre (CISTER), Porto 4249-015, Portugal, and also with CyLab Security and Privacy Institute, Carnegie Mellon University, Pittsburgh, PA 15213, USA (E-mail: kaili@ieee.org).}
\thanks{B.P.L.~Lau is with Singapore University of Technology and Design, SG 487372, Singapore (E-mail: billy\_lau@sutd.edu.sg).}
\thanks{X.~Yuan and W.~Ni is with the Digital Productivity and Services Flagship, Commonwealth Scientific and Industrial Research Organization (CSIRO), Sydney, NSW 2122, Australia (E-mail: \{xin.yuan, wei.ni\}@data61.csiro.au).}
\thanks{M.~Guizani is with the Machine Learning Department, Mohamed Bin Zayed University of Artificial Intelligence (MBZUAI), Abu Dhabi, UAE (E-mail: mguizani@ieee.org).}
\thanks{C.~Yuen is with Nanyang Technological University, SG 639798, Singapore (E-mail: chau.yuen@ntu.edu.sg).}
\thanks{Copyright (c) 20xx IEEE. Personal use of this material is permitted. However, permission to use this material for any other purposes must be obtained from the IEEE by sending a request to pubs-permissions@ieee.org.}
}

\markboth{}%
{Li \MakeLowercase{\textit{et al.}}: Towards Ubiquitous Semantic Metaverse: Challenges, Approaches, and Opportunities}

\IEEEcompsoctitleabstractindextext{%
\begin{abstract}
\boldmath In recent years, ubiquitous semantic Metaverse has been studied to revolutionize immersive cyber-virtual experiences for augmented reality (AR) and virtual reality (VR) users, which leverages advanced semantic understanding and representation to enable seamless, context-aware interactions within mixed-reality environments. This survey focuses on the intelligence and spatio-temporal characteristics of four fundamental system components in ubiquitous semantic Metaverse, i.e., artificial intelligence (AI), spatio-temporal data representation (STDR), semantic Internet of Things (SIoT), and semantic-enhanced digital twin (SDT). We thoroughly survey the representative techniques of the four fundamental system components that enable intelligent, personalized, and context-aware interactions with typical use cases of the ubiquitous semantic Metaverse, such as remote education, work and collaboration, entertainment and socialization, healthcare, and e-commerce marketing. 
Furthermore, we outline the opportunities for constructing the future ubiquitous semantic Metaverse, including scalability and interoperability, privacy and security, performance measurement and standardization, as well as ethical considerations and responsible AI. Addressing those challenges is important for creating a robust, secure, and ethically sound system environment that offers engaging immersive experiences for the users and AR/VR applications. 
\end{abstract}

}

\maketitle

\IEEEdisplaynotcompsoctitleabstractindextext
\IEEEpeerreviewmaketitle

\section{Introduction}
\label{sec_introduction}

\begin{figure*}[h]
	\centering
	\includegraphics[width=0.9\textwidth]{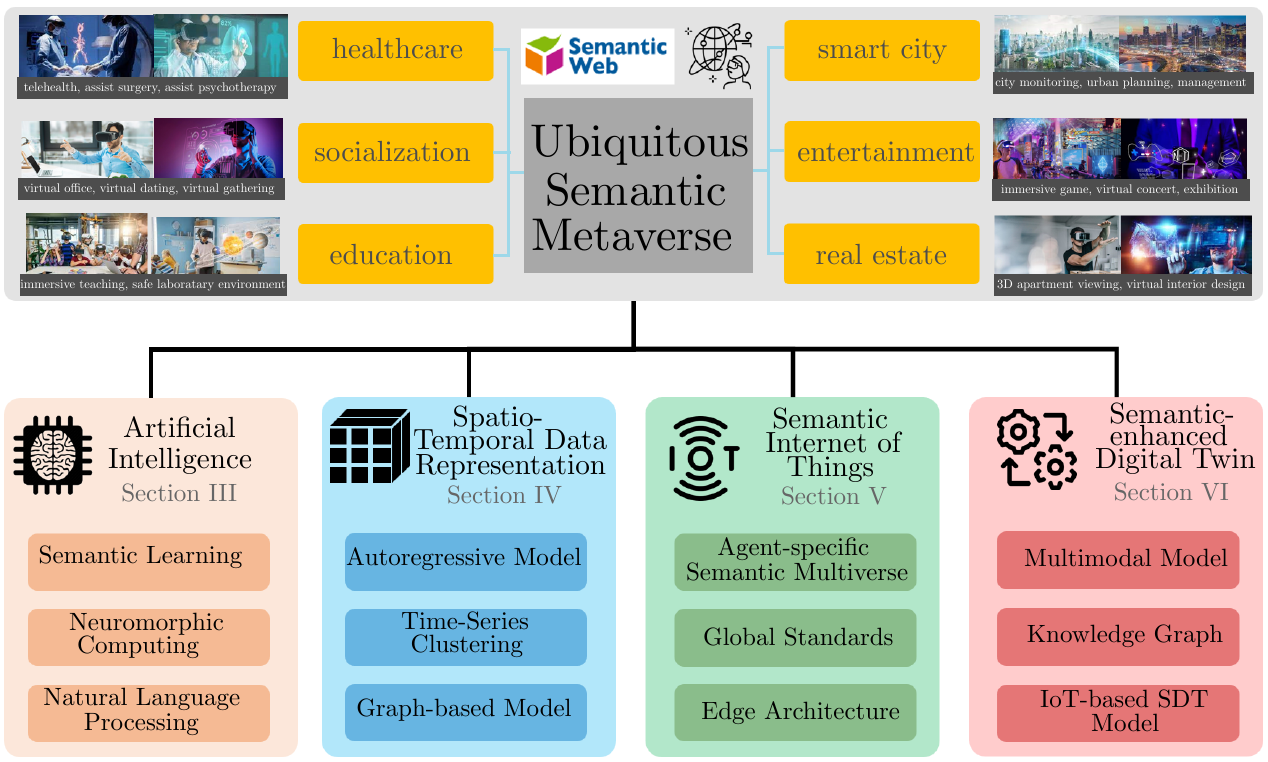} 
	\caption{An overview of fundamental components in ubiquitous semantic Metaverse regarding intelligence and spatio-temporal characteristic.}
	\label{fig:overallView}
\end{figure*}

Metaverse is widely regarded as the future of the internet, as evidenced by the surge in interest in late 2021, with a peak score of 100 on Google's search analytics~\cite{hackl2022navigating}. This interest has remained high throughout 2022, leading analysts to predict that Metaverse will be a top tech trend in 2023 and beyond. The objective of an ubiquitous Metaverse is to provide users with an anywhere-and-anytime immersive virtual experience using real-time interactive applications like augmented reality (AR) and virtual reality (VR)~\cite{tian2021scanning,tarasenko2021using}. The ubiquitous Metaverse has the potential to offer a collaborative and decentralized brand experience to users in various industries and services. For example, users equipped with AR/VR devices at home can enter a virtual store in their avatar form and access real-world inventory that can be selected and interacted with in the virtual world; a Metaverse-based system~\cite{gu2023metaverse} can be designed for emergency evacuation to ensure life safety. 

The ubiquitous Metaverse is an environment where the boundaries between the physical and digital spaces are blurred, enabled by the integration of numerous AR/VR devices and wireless communication networks~\cite{cheng2022will}. This convergence allows for the exchange of vast amounts of real-time 3D video content among the AR/VR devices~\cite{kutt2017semantically}. This increased data transfer results in conventional communication systems like 4G and 5G approaching the Shannon limit in the near future, with the remaining available spectrum becoming scarce. To address this bandwidth bottleneck in the Metaverse, next-generation communication systems must be capable of transmitting large volumes of AR/VR data.

In recent years, multitudinous ubiquitous semantic Metaverse technologies have attracted more and more researchers' attention~\cite{ismail2022semantic,ng2022stochastic}. \textit{The ubiquitous semantic Metaverse refers to a virtual, interconnected space that incorporates artificial intelligence (AI) and a comprehensive understanding of spatio-temporal and semantic information. The ubiquitous semantic Metaverse can enable seamless navigation, interaction, and anywhere-and-anytime immersive experiences within AR/VR environments across various platforms and devices}~\cite{kountouris2021semantics}. Moreover, the ubiquitous semantic Metaverse enables AR/VR applications to interpret human language and meaning, making it possible for people and hardware equipment to communicate effectively. This is significantly useful for tasks that require large-scale, real-time data transmissions and AI-aided analysis, such as information retrieval, text mining, natural language processing (NLP), and knowledge representation~\cite{manolova2021context,grobelnik2009semantic}. 

For instance, the ubiquitous semantic Metaverse can create a holistic understanding of a smart city's environment. By processing the spatial (location-based) and temporal (time-based) aspects of the data (based on traffic cameras, air quality sensors, and energy meters), city planners can identify patterns, such as congested areas, periods of poor air quality, and times of high energy consumption. The semantic aspect of the Metaverse enables the system to understand the relationships and contextual information related to the data, such as the impact of weather conditions on traffic flow or the correlation between air quality and energy consumption patterns. Using the insights gained from this analysis, city planners can make informed decisions on infrastructure improvements, traffic management strategies, and sustainable energy policies.

Rather than transmitting raw data, the ubiquitous semantic Metaverse can enable the sender and receiver devices to utilize AI techniques to understand the meaning and context of the real-time data being exchanged~\cite{wang2023semantic,maatouk2022age}. This allows devices to optimize the transmission and processing of the data, which can result in communication bandwidth savings. For instance, an AR/VR device can use the ubiquitous semantic Metaverse to annotate the data with AI metadata that describes the data's meaning and context, such as its type, level of importance, or intended audience~\cite{lu2022rethinking}. The AI metadata is then sent to a data center that can process and utilize the data effectively. In addition, the ubiquitous semantic Metaverse can facilitate a shared understanding between the sender and receiver by using a common ontology, which is an AI-generated data model representing the meaning of the data. Using a common set of terms and concepts, the sender and receiver devices can describe the data in a mutually understandable way.

In this paper, we investigate the intelligence and spatio-temporal characteristics of four fundamental components in the ubiquitous semantic Metaverse, i.e., artificial intelligence (AI), spatio-temporal data representation (STDR), semantic Internet of Things (SIoT), and semantic-enhanced digital twin (SDT); see Figure~\ref{fig:overallView}. This figure provides an overall view of the relationships and interactions between these components, highlighting their key roles in enabling intelligent behavior and spatio-temporal reasoning in the virtual environment of the ubiquitous semantic Metaverse. Specifically, 
\begin{itemize}
    \item AI technologies, such as NLP, computer vision, and machine learning~\cite{lim2022realizing}, can enable the ubiquitous semantic Metaverse to understand and process human language, recognize objects and scenes, and learn from data to make predictions and improve over time. These capabilities allow the ubiquitous semantic Metaverse to provide natural, engaging, and personalized experiences for users. 
    \item STDR allows the ubiquitous semantic Metaverse to model and process information concerning the location and time of events, objects, and users~\cite{cressie2015statistics}. This enables an accurate and context-aware understanding of user interactions, leading to immersive and realistic experiences. Spatio-temporal data also supports advanced analytics and decision-making processes, such as predicting user behavior and optimizing content delivery. 
    
    \item SIoT is a technological paradigm that combines the concepts of IoT and the semantic web to create an intelligent and interconnected network of IoT devices and data~\cite{rhayem2020semantic}. SIoT can provide a semantic network structure for AR/VR applications based on IoT. SIoT aims to enhance the discoverability and reusability of IoT devices and data by introducing a semantic layer that adds meaning and context to the information exchanged among connected devices~\cite{palavalli2016semantic}. This semantic layer enables devices to communicate and cooperate with each other, even if they use different protocols or data formats. The semantic layer can establish a common language that facilitates the integration and automation of complex tasks~\cite{kovacs2016standards}. Furthermore, the SIoT has the potential to revolutionize various fields, such as healthcare, energy management, transportation, and smart cities, by providing more efficient, secure, and personalized services.

    \item SDT can be developed to improve immersive cyber-virtual experiences by allowing natural and intuitive interactions between the AR/VR devices and cyber-physical environments~\cite{joda2022internet}. This can be achieved through NLP and understanding, which helps virtual agents and objects respond to the user input of the AR/VR device in a meaningful way. SDT can also allow more contextually relevant information to be presented in the virtual environment. This can include things like location-based information, personalized content, and real-time updates. In addition, the use of SDT in Metaverse can lead to more immersive and engaging experiences for the users. 
\end{itemize}

\begin{figure}[h]
	\centering
	\includegraphics[width=0.4\textwidth]{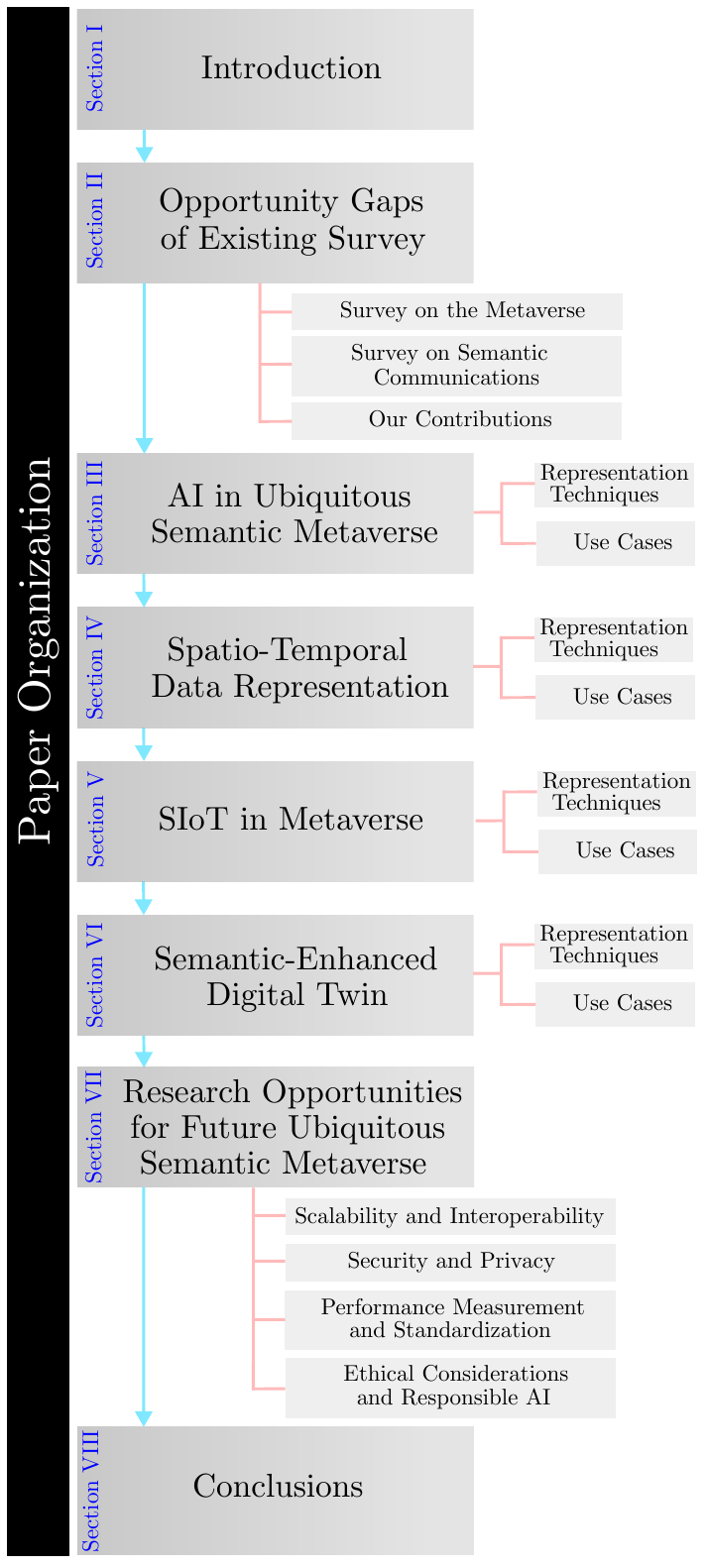} 
	\caption{The paper organization shows the structure of this survey.}
	\label{fig:organization}
\end{figure}

Consider the ubiquitous semantic Metaverse for smart city management, where real-time data from various SIoT devices is used to create a seamless and efficient urban environment. Specifically, a large number of the SIoT devices can be deployed, including traffic sensors, air quality monitors, and energy meters. SIoT enables these devices to communicate and exchange data in a meaningful, interoperable manner. The data collected from these SIoT devices is inherently spatio-temporal, associated with a specific location and time. STDR allows efficient representation and processing of this data, enabling the system to understand trends and patterns over space and time. 

AI is applied to analyze the data gathered and exchanged through the SIoT and structured by STDR. Machine learning models can predict future patterns and make data-driven decisions. This can allow the city management to proactively manage traffic flow, and could optimize energy usage across the city based on predicted demand. In addition, for each physical entity (such as a building, a street, or even the entire city), an SDT can be created in the ubiquitous semantic Metaverse. These digital twins mirror the real-time state of their physical counterparts using the data from the SIoT devices. Changes in the physical world (such as traffic congestion or energy usage spikes) are reflected in the SDT in support of fast and effective city management in the ubiquitous semantic Metaverse.

We also investigate the opportunities and future research directions for constructing the ubiquitous semantic Metaverse, including scalability and interoperability, privacy and security, performance measurement and standardization, ethical considerations and responsible AI. Addressing those challenges is vital for creating a robust, secure, and ethically sound environment that offers engaging, immersive experiences for the users. Tackling issues related to scalability and interoperability ensures seamless integration and interaction between various platforms, devices, and data formats. Addressing privacy and security concerns protects user data and maintains trust in the ubiquitous semantic Metaverse. Establishing performance measurement and standardization helps create consistent evaluation metrics, benchmarks, and best practices, facilitating the interoperability and reliability of ubiquitous semantic Metaverse technologies. In addition, ethical considerations and responsible AI ensures that the ubiquitous semantic Metaverse development adheres to ethical principles, avoids biases, and promotes inclusivity and accessibility for diverse user groups.

Figure~\ref{fig:organization} presents a paper organization for visualizing the structure. The rest of this survey is organized as follows. In Section~\ref{sec_literature}, we examine the gaps in existing surveys pertaining to Metaverse, semantic communications, as well as SIoT. Section~\ref{sec_AI} studies AI technologies that enable the ubiquitous semantic Metaverse to process human language, recognize objects, and learn from data for efficient user engagement. Section~\ref{sec_STDR} studies STDR that allows the ubiquitous semantic Metaverse to model location and time information for immersive and realistic experiences. Section~\ref{sec_SIOT} presents SIoT that integrates semantic communications with IoT devices and systems, providing seamless interaction between real-world data and virtual environments. Section~\ref{sec_SDT} discusses SDT that creates virtual representations of physical objects, systems, or processes, enabling advanced simulations, analysis, and optimization within the ubiquitous semantic Metaverse. The research opportunities for building future ubiquitous semantic Metaverse are delineated in Section~\ref{sec_future}. Section~\ref{sec_conclusions} concludes the survey. 

\section{Opportunity Gaps of Existing Surveys}
\label{sec_literature}
In this section, we review the literature thoroughly in terms of surveys on the Metaverse and semantic communications. 

\begin{table*}[h]
\caption{Summary of existing surveys on Metaverse and semantic communications}
\label{tbl:Comm}
\begin{center}
\begin{tabular}{@{}c|l|l|c|c|c|c@{}}
\toprule
Literature & Descriptions \& Contributions  & Review Genres & Metaverse & \begin{tabular}[c]{@{}c@{}} Semantic \\ Communication \end{tabular}& AI & \begin{tabular}[c]{@{}c@{}} Spatio-temporal \\ characteristics \end{tabular}\\ \midrule
\cite{wang2022survey,sun2022metaverse} & 
\begin{tabular}[c]{@{}l@{}}Review on the current state metaverse \\ and
talks about the privacy concern \\ within the applications of Metaverse.\end{tabular}&
\begin{tabular}[c]{@{}l@{}} Generic review and analysis\end{tabular} & \checkmark & $\times$ & $\times$ & $\times$ \\ \midrule
\cite{ning2021survey} & \begin{tabular}[c]{@{}l@{}}Analyze on infrastructure for enabling \\ Metaverse technology and integration \\ methodology.\end{tabular} &  
\begin{tabular}[c]{@{}l@{}} Infrastructure development\end{tabular}  & \checkmark & $\times$ & $\times$ & $\times$ \\ \midrule
\cite{zhang2022parallel} & \begin{tabular}[c]{@{}l@{}} Integrate the Metaverse and transportation \\ technology for improving the system's \\intelligence.\end{tabular} &  
\begin{tabular}[c]{@{}l@{}} Intelligent transportation systems\end{tabular}  & \checkmark & $\times$ & $\times$ & $\times$ \\ \midrule
\cite{chengoden2023metaverse} &  \begin{tabular}[c]{@{}l@{}} Discuss about challenges and integration \\ of Metaverse technology in health services. \end{tabular} &  Health  & $\checkmark$ & $\times$ & $\times$ & $\times$ \\ \midrule
\cite{fu2022survey, huang2022fusion} &  \begin{tabular}[c]{@{}l@{}} Investigate the technological challenges  \\in blockchain and integration of it in \\Metaverse. \end{tabular} &  Blockchain  & $\checkmark$ & $\times$ & $\times$ & $\times$ \\ \midrule
\cite{huynh2023artificial} &  \begin{tabular}[c]{@{}l@{}} Inspect the development of AI and its \\ realization of its various platforms. \end{tabular} &  Smart cities & $\times$ & $\times$ & \checkmark & $\times$ \\ \midrule
\cite{huang2023security} &  \begin{tabular}[c]{@{}l@{}} Introduce four elements to review current \\ state of Metaverse security and privacy. \end{tabular} & \begin{tabular}[c]{@{}l@{}} Security and privacy\end{tabular} & \checkmark
& $\times$ & $\times$ & $\times$\\ \midrule
\cite{li2022internet} &  \begin{tabular}[c]{@{}l@{}} Examine four key technologies for  \\enabling AR/VR in IoT-inspired Metaverse. \end{tabular} & \begin{tabular}[c]{@{}l@{}} Integration design\end{tabular} & \checkmark & $\times$ & $\times$ & $\times$ \\ \midrule
\cite{yang2022semantic} & 
\begin{tabular}[c]{@{}l@{}} Study the integration of semantic \\ communication and its applications. \end{tabular}&
\begin{tabular}[c]{@{}l@{}} Generic review and analysis \end{tabular} & $\times$ & \checkmark & $\times$ & $\times$ \\ \midrule
\cite{luo2022semantic, lan2021semantic, qin2021semantic} & 
\begin{tabular}[c]{@{}l@{}} Review the usage of AI in performing\\ end-to-end communication based \\ semantic communications.\end{tabular}&
\begin{tabular}[c]{@{}l@{}} Integration design \end{tabular} & $\times$ & \checkmark & \checkmark & $\times$ \\ \midrule
\cite{iyer2022survey} & 
\begin{tabular}[c]{@{}l@{}} Provide insights on transmission model \\ improvement through channel coding to \\ improve accuracy of transmitted information. \end{tabular}&
\begin{tabular}[c]{@{}l@{}} Transmission model design \end{tabular} & $\times$ & \checkmark & $\times$ & $\times$  \\ \midrule
\cite{wheeler2022engineering} & 
\begin{tabular}[c]{@{}l@{}} Propose a knowledge graph for efficient \\and accurate responses on semantic \\communications. \end{tabular}&
\begin{tabular}[c]{@{}l@{}} Structure representation design \end{tabular}  & $\times$ & \checkmark & $\times$ & $\times$ 
 \\ \midrule
\cite{lu2022semantics} & 
\begin{tabular}[c]{@{}l@{}} Talks about the enabling technique to \\accurately interpret message in semantic \\communication. \end{tabular}&
\begin{tabular}[c]{@{}l@{}} Language model design \end{tabular}  & $\times$ & \checkmark & $\times$ & $\times$  \\ \midrule
\cite{yang2023secure} & 
\begin{tabular}[c]{@{}l@{}} Share the techniques to ensure secure \\semantic communication in various aspects.\end{tabular}&
\begin{tabular}[c]{@{}l@{}} Secure communication design \end{tabular}  & $\times$ & \checkmark & $\times$ & $\times$ \\ \midrule
\cite{du2022rethinking, rahman2020comprehensive} & 
\begin{tabular}[c]{@{}l@{}} Highlight the techniques to enhance \\ communication security when applied to IoT.\end{tabular}&
\begin{tabular}[c]{@{}l@{}} IoT communication design \end{tabular}  & $\times$ & \checkmark & $\times$ & $\times$ \\ \midrule
\textbf{This survey} & 
\begin{tabular}[c]{@{}l@{}} Investigate representative techniques and typical \\ use cases regarding the four fundamental \\ components in ubiquitous semantic Metaverse. \\ Outline the opportunities of building future \\ubiquitous semantic Metaverse.\end{tabular} &
\begin{tabular}[c]{@{}l@{}} Spatio-temporal characteristics \\ of AI, STDR, SIoT and SDT \end{tabular} & \checkmark & \checkmark & \checkmark & \checkmark \\  \bottomrule
\end{tabular}
\end{center}
\end{table*}

\subsection{Existing Surveys on Metaverse Techniques and Applications}
In this section, we present existing surveys regarding Metaverse in recent years. A summarized list of surveys is consolidated in Table~\ref{tbl:Comm}. In~\cite{wang2022survey}, the authors review the concepts and applications of the Metaverse, including virtual gaming, social networking, education, and ecommerce. Then, they present the architecture, networking, virtual environments, user interfaces, and security threats in the Metaverse.

The components of the current state of the Metaverse industry, including AR/VR, major players and their business models, are introduced in~\cite{sun2022metaverse}. The players can be the potential applications of the Metaverse used for education, healthcare, entertainment, and social interaction. The Metaverse industry also has the security and privacy concerns, such as identity theft, cyber-attacks, and data breaches, and potential solutions are discussed in~\cite{sun2022metaverse}.

The authors of~\cite{ning2021survey} focus on the development status of the Metaverse with regard to network infrastructure, management technology, basic standard technology, VR object connection and convergence. The Metaverse integrates multi-technology dominance, sociality, and hyperspace characteristics for building immersive 3D virtual social worlds. In~\cite{zhang2022parallel}, the Metaverse can be utilized in intelligent transportation systems for reducing traffic accidents and improving driving safety, where digital models are built to simulate the virtual transportation space. The authors present several exist various significant hurdles in perceiving the environment of the Metaverse, including the applicability of vision techniques to different traffic scenarios, and the integration of data from multiple sensors.

The Metaverse can enable healthcare professionals to provide remote care to patients, virtual medical consultations, medical education and training, and patient engagement~\cite{chengoden2023metaverse}. The challenges associated with the integration of Metaverse technology in healthcare are examined, including data privacy and security, regulatory compliance, ethical concerns, and technical barriers. In~\cite{fu2022survey}, blockchain and intelligent networking are studied in the Metaverse applications to improve security, privacy, and interoperability. Blockchain can be used for decentralized identity management, digital asset management, and smart contracts in the Metaverse. In~\cite{huynh2023artificial}, the authors examine several deep learning and machine learning architectures used in the Metaverse. They concentrate on AI-based approaches to several technological areas that can potentially create virtual worlds in the Metaverse. 

The authors of~\cite{huang2023security} examine identity theft, virtual asset theft, cyberbullying, and privacy invasion in the Metaverse, and study the limitations of existing encryption, authentication, and access control solutions. For improving privacy, blockchain-based smart contracts can create a decentralized system for managing virtual assets, while enforcing rules and regulations in the Metaverse, such as access control and privacy policies. The applications of building information modeling (BIM) and blockchain in the Metaverse are presented in~\cite{huang2022fusion}, where BIM is a digital representation of the physical and functional characteristics of a building or construction project. Blockchain-based smart contracts can be used with BIM in the Metaverse to automate payment processes and ensure the quality of virtual building components.

In our earlier work~\cite{li2022internet}, the convergence of the IoT and Metaverse technologies is discussed, potentially creating an immersive and interconnected virtual world that combines physical and digital experiences. Several use cases of the IoT-Metaverse convergence, such as smart cities, healthcare, and gaming, are presented. The IoT provides real-time data about the physical world, enhancing the virtual world in the Metaverse. Moreover, an insightful discussion was given to highlight the importance of addressing data privacy and security issues and the need for new standards and protocols to enable interoperability between IoT and Metaverse technologies.

\subsection{Existing Surveys on Semantic Communications Techniques and Applications}
In this section, we have compiled a list of existing surveys about semantic communication as shown in Table~\ref{tbl:Comm}. 
Semantic communications can be used to improve the effectiveness of communication by providing additional context and information to help convey meaning and understanding~\cite{yang2022semantic}. Semantic communications recognize that communication is more than just the transmission of information and that meaning and interpretation are critical components of effective communication. Semantic communication is for the recognition that different people have different ways of expressing themselves and interpreting messages. Semantic communications provide a framework for accommodating these individual differences by allowing people to use language, context, and social cues to convey meaning in a way that is tailored to their needs. 

In~\cite{luo2022semantic}, an overview of the feature extraction-based semantic communications is presented. As semantic communications are different from conventional communications in terms of wireless channels, source and channel encoding (decoding), and performance metrics, the design of the semantic system has to be adapted to the types of messages transmitted. Thus, the AI structures of the source encoder (decoder) and channel encoder (decoder) can be very much different. 

By incorporating semantic communications into transmission models as well as source and channel coding, information errors can be reduced, and the accuracy of information transmission is improved~\cite{iyer2022survey}. In source coding, semantic communications can identify the most relevant and important data to be transmitted. This reduces the amount of data that needs to be transmitted, improving the efficiency of the system. In channel coding, semantic communications improve the accuracy of the encoding.

In~\cite{lan2021semantic}, it presents that semantic communications provide benefits for three types of communication, i.e., human-to-human (H2H), human-to-machine (H2M), and machine-to-machine (M2M). For H2H, semantic communications allow individuals to transmit and interpret messages based on the intended meaning, which increases understanding between individuals. For H2M, semantic communications can enhance the ability of machines to understand and respond to human requests or commands. For M2M, semantic communications enable efficient data exchange between machines.

In~\cite{wheeler2022engineering}, knowledge graphs are studied to provide a structured representation of the knowledge that can be shared in semantic communications and understood by the sender and receiver. This could be particularly useful in situations where different systems use different terminologies or vocabularies. Moreover, deep learning models can be trained to learn the relationships between words and phrases in semantic communications and use that knowledge to interpret messages and provide more accurate responses. 

The authors in~\cite{lu2022semantics} present enabling techniques, such as NLP and machine learning, which identify the meaning of words and phrases in context, enabling accurate interpretation of messages in semantic communications. Implicit and explicit reasoning can also be applied in semantic communications. For example, in a semantic medical diagnosis, a doctor may use implicit reasoning to interpret a patient's symptoms and medical history while using explicit reasoning to make a diagnosis based on that information.

An overview of the current state of communication systems is offered in~\cite{qin2021semantic}, where the limitations of traditional communication systems are discussed. As an advanced technique, semantic communications are expected to enable more accurate and effective communications than traditional ones. The principles of semantic communications are outlined, i.e., the knowledge graphs, deep learning, and reasoning-based methods, which can be utilized in healthcare, finance, and transportation. 

Since semantic communications transmit and interpret messages based on their meaning, which may contain sensitive or valuable information, attackers can target the meaning of the message, in addition to the syntax or structure of the message~\cite{yang2023secure}. Moreover, most semantic communications are generated using machine learning (ML)-based methods. Attackers can also attempt to manipulate or corrupt the learning models to obtain access to sensitive information. Therefore, the security challenges in semantic communications require new approaches to security design, such as defense algorithms, protocols, and mechanisms to protect the meaning of the message and ML used to generate semantic information. 

\begin{table}[htb]
\centering
\caption{An overview of Sections~\ref{sec_AI} --~\ref{sec_SDT}}
\label{tb_section}
\begin{tabularx}{8cm}{l|X}
\hline
\bf{The four components} & \bf{Highlights} \\ \hline
Section~\ref{sec_AI}: AI & Representative techniques regarding semantic learning, neuromorphic computing, and natural language processing; Use cases in medical ubiquitous semantic Metaverse, virtual assistants, and educational training. \\ 
& \\
Section~\ref{sec_STDR}: STDR & Representative techniques regarding spatio-temporal autoregressive models, time-series clustering, and graph-based models; Use cases in smart health, environmental simulations, and virtual collaborative activities. \\
&  \\
Section~\ref{sec_SIOT}: SIoT & Representative techniques regarding agent-specific semantic multiverses, global standards, and federated edge architectures; Use cases in  smart home systems, smart cities, and healthcare. \\
& \\
Section~\ref{sec_SDT}: SDT & Representative techniques regarding multimodal modes, knowledge graphs, and IoT-based models; Use cases in urban planning, building's energy-consuming systems, and industrial manufacturing. \\
& \\
\hline
\end{tabularx}
\end{table}

The conventional security methods used for bit transmission in wireless communication cannot be directly implemented in SIoT, which emphasizes semantic information transmission. Thus, wireless communication security has to be reconsidered in SIoT~\cite{du2022rethinking}. The characteristics of physical layer security, covert communications, and encryption in SIoT are presented. At the same time, secrecy outage probability and detection failure probability can be used as two performance indicators to measure SIoT's security. In~\cite{rahman2020comprehensive}, it is claimed that the inter-operation among the heterogeneous IoT components is essential for providing semantic interoperability to SIoT. Given various system requirements, interoperability in SIoT can be applied on different levels, such as technical, syntactic, semantic, and organizational interoperability.

\begin{figure*}[h]
	\centering
	\includegraphics[width=0.9\textwidth]{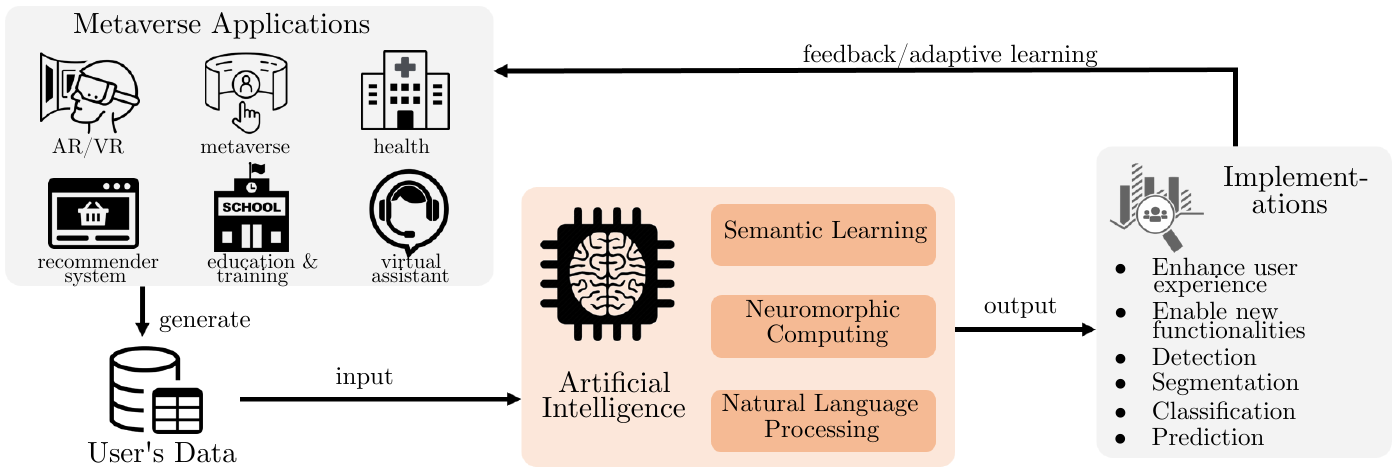} 
	\caption{The applications of AI in the ubiquitous semantic Metaverse, including AR/VR, intelligent health, recommender system, remote education and training, and virtual assistants, which enable the Metaverse to provide more natural, engaging, and personalized experiences for users.}
	\label{fig:ai}
\end{figure*}

\subsection{Our Contributions}
The existing surveys and tutorials focus on either the applications in Metaverse or the human-machine interaction in semantic communications, with an aim to improve users' virtual experience or communication efficiency. However, the ubiquitous semantic Metaverse that incorporates AI and a comprehensive understanding of spatio-temporal and semantic information is still unexplored. The key contributions of the paper are summarized, as follows. 
\begin{itemize}
\item This paper comprehensively surveys the representative techniques and use cases of four fundamental components (i.e., AI, STDR, SIoT, and SDT) in the ubiquitous semantic Metaverse, as outlined in Table~\ref{tb_section}. The four components provide users with an anywhere-and-anytime immersive experience, and are essential to addressing critical research and development challenges of the ubiquitous semantic Metaverse. 

\item We focus on the intelligence and spatio-temporal characteristics of the four fundamental components. We study that ensuring seamless, intuitive interactions between diverse devices, platforms, and data formats can be a daunting task due to a large, complex space with countless users, objects, and interactions. Not only does the Metaverse need to respond in real-time to changes in both the virtual and real worlds, but also guarantees that the user's virtual experiences are personalized based on its behavior and preferences. 

\item We also investigate the opportunities and future research directions for constructing the ubiquitous semantic Metaverse, including scalability and interoperability, privacy and security, performance measurement and standardization, ethical considerations, and responsible AI. Addressing those challenges is vital for creating a robust, secure, and ethically sound environment that offers engaging, immersive experiences for the users.
\end{itemize}

\begin{figure*}[h]
	\centering
	\includegraphics[width=0.97\textwidth]{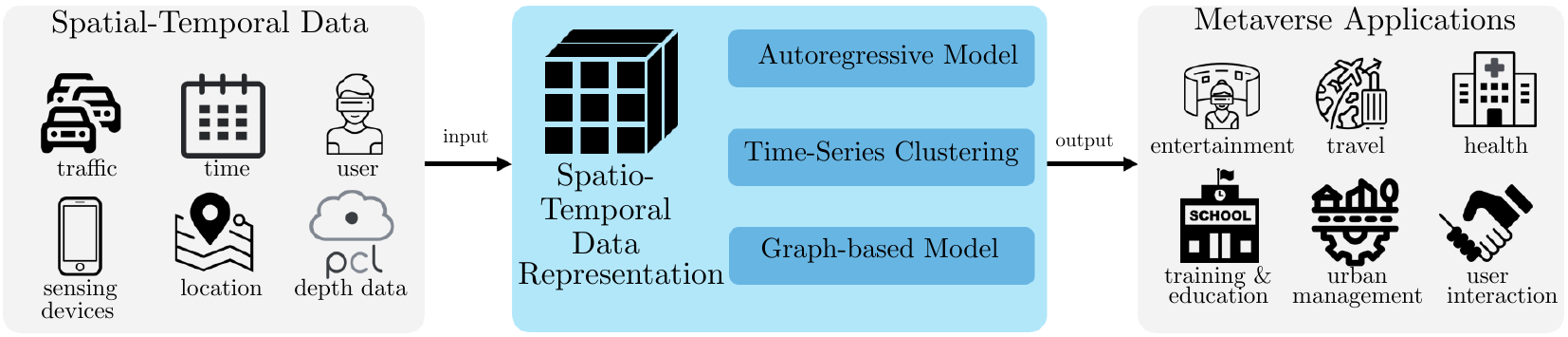} 
	\caption{A breakdown of the typical techniques used in STDR, including spatio-temporal autoregressive models, time-series clustering, and graph-based models.}
	\label{fig:stdr}
\end{figure*}

\section{AI in Ubiquitous Semantic Metaverse}
\label{sec_AI}
In this section, we specifically investigate the representative techniques and use cases of the AI technologies that enable the ubiquitous semantic Metaverse to process human language, recognize objects, and learn from data for better user engagement.

\subsection{Representative techniques} 
\subsubsection{Problems and gaps} AI technologies, such as semantic learning, neuromorphic computing, and natural language processing (NLP), can enable ubiquitous semantic Metaverse to understand human languages. The AI technologies can also recognize objects and scenes, and learn from spatio-temporal data to make predictions and improve over time. This can help improve the accuracy of predictions and decision-making, leading to more efficient and effective systems. Figure~\ref{fig:ai} illustrates the applications of AI in the ubiquitous semantic Metaverse, including AR/VR, intelligent health, recommender system, remote education and training, and virtual assistants, which enable the Metaverse to provide more natural, engaging, and personalized experiences for users. These AI capabilities allow for seamless interaction with virtual agents, more realistic perception and manipulation of virtual objects. Additionally, the AI technologies improves automation of virtual systems, which enhances the overall anywhere-and-anytime immersive experience in the ubiquitous semantic Metaverse.

\subsubsection{Impacts} In~\cite{zeng2022hfedms}, the applications in ubiquitous semantic Metaverse suffer from non-independent and identically distributed (non-i.i.d.) data and learning degradation due to the streaming data and limited communication bandwidth. A dynamic sequential-to-parallel training-based federated learning is developed with semantic compression and compensation to fuse compressed historical data semantics and calibrate classifier parameters. In the ubiquitous semantic Metaverse, the meaning of information is defined by the receiver's practical task and remains undisclosed to the sender. The genuine data observed by the sender may display a dissimilar distribution to the data located in the communal background knowledge archive. To address this issue, neural network-based semantic communication systems can be developed~\cite{zhang2022deep,xiao2022reasoning,wang2023multimodal} for image transmission, in which the transmitter is not aware of the task, and the data environment continually changes. For speech recognition and synthesis, deep learning-driven semantic communication systems~\cite{weng2022deep,han2022semantic,xie2021task} are designed for speech transmission. By utilizing a semantic transmitter based on Convolutional Neural Networks (CNN) and Recurrent Neural Networks (RNN), text-related semantic features can be extracted from the input speech, which reduces the amount of data transmitted and the necessary communication resources~\cite{weng2021semantic,weng2021semantic2}. 

In~\cite{chen2023neuromorphic} and~\cite{zhou2021semantic}, neuromorphic computing is integrated with semantic communication to maintain energy consumption levels across the sensing-processing-communication sequence, adapting to the changing conditions of the monitored environment. A comprehensive design of the neuromorphic computing framework is also showcased, utilizing supervised learning through surrogate gradient descent techniques. Empowered by deep learning, an NLP-based semantic communication system is created for text transmission~\cite{xie2020deep,mu2022semantic}, which can maximize the system capacity while minimizing the semantic errors by restoring the meaning of sentences, as opposed to addressing bit or symbol errors typically encountered in traditional communication systems. 

AI algorithms can be used with ubiquitous computing to optimize the usage of resources in resource-limited devices in ubiquitous semantic Metaverse~\cite{baccour2022pervasive}. This can prioritize tasks for immersive experiences based on various parameters, such as user demands, data requirements, latency tolerance, and energy usage. AI algorithms adapt based on the user's demands and time-varying ubiquitous computing environments, while reducing the computing complexity without affecting the training accuracy.

\subsection{Use cases} AI techniques can be applied to various use cases in ubiquitous semantic Metaverse to enhance the user experience and enable new functionalities. For example, AI can be integrated with medical technologies in ubiquitous semantic Metaverse to aid in creating, testing, assessing, regulating, implementing, and improving AI-driven medical practices, particularly those related to diagnostic and therapeutic medical imaging~\cite{wang2022development}. A number of applications can demonstrate the potential of medical ubiquitous semantic Metaverse with AI, such as medical imaging~\cite{huang2022toward}, virtual comparative scans, sharing of raw data, enhanced regulatory science, and ubiquitous semantic Metaverse-based medical interventions. 

AI-powered virtual assistants can help users navigate the ubiquitous semantic Metaverse, answer questions, and perform tasks~\cite{yang2022semantic2,luong2022edge,ruminski2018architecture}. By leveraging NLP, knowledge graphs, and recommender systems, these assistants can provide personalized guidance and support to users based on their preferences and interests. 
AI techniques can also facilitate social interactions and collaborations in the ubiquitous semantic Metaverse. NLP-driven chatbots and dialogue systems can simulate realistic conversations~\cite{peng2022robust}, while computer vision can enable expressive avatars that mimic users' facial expressions and body language. Multi-agent systems can create a more dynamic environment where AI agents and users interact and collaborate~\cite{lu2021reinforcement}. 

In addition, AI techniques can be used in educational and training contexts to create adaptive learning experiences tailored to individual users~\cite{hare2022hierarchical,chaccour2022disentangling,li2022intelligent}. By leveraging recommender systems, personalized learning paths can be generated based on users' skills, knowledge, and preferences. In particular, NLP can be utilized to provide natural language explanations and feedback, while simulation and planning techniques can create realistic training scenarios.

\section{Spatio-Temporal Data Representation}
\label{sec_STDR}
In this section, we specifically investigate the representative techniques and use cases of STDR that allows the ubiquitous semantic Metaverse to model location and time information for more immersive and realistic experiences.

\subsection{Representative techniques} 
\subsubsection{Problems and gaps} STDR is a critical component of the ubiquitous semantic Metaverse that enables the accurate modeling and processing of information related to the location and time of events, objects, and users, leading to more immersive and realistic virtual experiences. Spatio-temporal autoregressive models are used in STDR to capture the dependencies and relationships between spatio-temporal variables, allowing for the accurate prediction and modeling of spatio-temporal data. Time-series clustering is also used in STDR to group similar spatial units based on their temporal patterns. It identifies regions with common temporal trends or behaviors, allowing for the discovery of spatio-temporal relationships in the data. Graph-based models in STDR represent spatio-temporal data as a network of nodes (locations) and edges (connections). The graph-based models, such as Graph Neural Networks (GNNs) or Graph Attention Networks (GATs), are typically used to model complex spatio-temporal dependencies and patterns in the data. Figure~\ref{fig:stdr} provides a breakdown of these techniques used in STDR, including spatio-temporal autoregressive models, time-series clustering, and graph-based models. 

\subsubsection{Impacts} In~\cite{chen2018novel}, traffic flow prediction with uncertain data features is studied, where an autoregressive data representation is employed to investigate temporal and spatial properties of traffic flow. The traffic-based data representation can help capture the dependencies between different locations, such as the impact of traffic congestion at one location on the traffic flow in nearby areas. By modeling these spatial correlations, the autoregressive models can accurately predict how traffic flow changes in one area may affect others~\cite{salman2021data,jiang2020audio,chen2016learning}. The time-series clustering data representation can allow for the integration of real-time traffic data, such as sensor readings or crowdsourced information from connected vehicles or mobile devices~\cite{guo2020optimized,andreoletti2019network,liu2020dynamic}. This real-time data can be used to update traffic flow predictions dynamically, making them more responsive to current conditions and enabling better traffic management. 

The graph-based models in STDR are typically used to model spatial relationships between objects, users, or other elements in the ubiquitous semantic Metaverse. GNNs and GANs can represent these spatial relationships as graph structures, where nodes represent the elements and edges represent their spatial connections~\cite{zhang2021network,shengwen2023intelligent}. By applying graph-based convolution operations, GNNs and GANs can learn complex spatial patterns and dependencies, enabling better understanding and prediction of the environment's dynamics~\cite{zhu2022eyeqoe,ding2023gan}. To incorporate temporal information in GNNs or GANs, time can be treated as an additional dimension in the graph, connecting nodes representing the same entities at different time steps. Alternatively, recurrent units like gated recurrent units or Long Short-Term Memory (LSTM) cells can be integrated into GNNs or GANs to capture the temporal dependencies and dynamics of the spatio-temporal data~\cite{huang2020long}, enabling more accurate predictions and simulations in VR/AR environments. Moreover, GNNs and GANs can help upgrade VR/AR systems with interactive and responsive applications by predicting users' behavior, object movements, or environmental changes based on spatio-temporal data. By updating the graph structure in real-time~\cite{xu2020real}, GNNs and GANs can enable VR/AR systems to adapt and respond to users' actions, leading to an immersive and engaging experience. 

In addition, STDR crucially supports ubiquitous computing in the semantic Metaverse, which deals with the representation and processing of ubiquitous data that changes over space and time~\cite{breitbach2019context}. In particular, STDR selects the most relevant features from raw data, which reduces the size of data without losing essential information. This helps balance the data processing accuracy and complexity, which is particularly important for resource-limited devices.

\subsection{Use cases} The ubiquitous semantic Metaverse can incorporate spatio-temporal data to create virtual environments tailored to individuals' mental health and well-being needs. For example, an end-to-end RhythmNet~\cite{niu2019rhythmnet} is developed for estimating heart rate remotely from facial data. STDR is employed to encode heart rate signals from multiple regions of interest volumes as input. Subsequently, the SRTD formulates a convolutional network for heart rate estimation. The connections between adjacent heart rate measurements are learned from a video sequence using a gated recurrent unit, leading to efficient heart rate measurement. 
AR/VR therapy sessions with STDR~\cite{fernandez2019deterioration} can be conducted in calming or therapeutic environments that change over time to support patients' progress and help them manage stress, anxiety, or other mental health conditions.

In~\cite{kellner2016tracking}, STDR is used to recreate historical or real-world locations, allowing users to visit and explore these places virtually. This includes experiencing different time periods or witnessing how the location changes over time, offering an immersive and educational experience. By incorporating spatio-temporal data, the ubiquitous semantic Metaverse can simulate various environmental conditions, such as changing weather patterns, day-night cycles, or natural disasters~\cite{jiang2020integrated,prasad2017parallel,wang2017spatial}. These simulations can be used for entertainment, education, or training purposes, such as disaster response or climate change awareness. STDR can be utilized in the metaverse to simulate real-world traffic conditions, public transportation systems, or urban infrastructure. This can help urban planners, traffic engineers, or policymakers to visualize and analyze the impact of different strategies or interventions, leading to better-informed decision-making.

In addition, STDR can help model and support real-time interactions between users, avatars, or objects in the ubiquitous semantic Metaverse. This enables collaborative activities, competitive events, or other synchronous experiences within the virtual environment, fostering social connections and shared experiences~\cite{chen2021exploring,song2021attention,nie2022real}. STDR can also create realistic and dynamic simulations for various training and educational purposes~\cite{hsieh2018preliminary}. For example, medical professionals can practice surgical procedures in a virtual environment that mimics real-life scenarios~\cite{mozumder2022overview,sugimoto2021extended,schott2021vr}, or emergency response teams can train in virtual disaster situations that evolve over time~\cite{zhou2021vr,chakareski2017aerial}.

\begin{figure*}[h]
	\centering
	\includegraphics[width=0.85\textwidth]{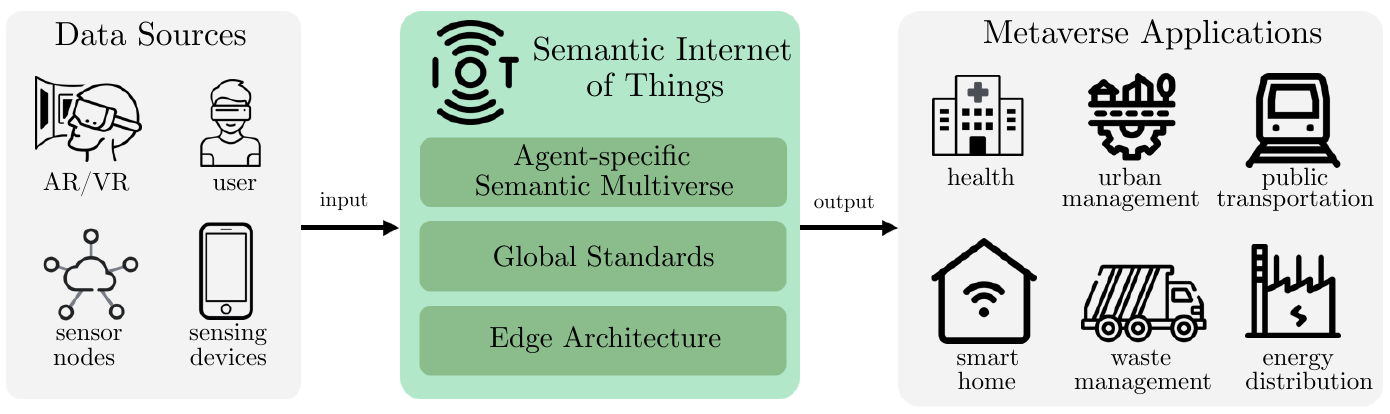} 
	\caption{The use of SIoT in addressing the challenges of IoT-based contextual information provisioning, where SIoT provides a semantic layer that allows devices to understand the meaning of the data they collect and share.}
	\label{fig:siot}
\end{figure*}

\section{SIoT in Metaverse}
\label{sec_SIOT}
In this section, we specifically investigate the representative techniques and use cases of SIoT. The techniques and use cases that integrate semantic communications with IoT devices and systems can provide seamless interaction between real-world data and virtual environments.

\subsection{Representative techniques} 
\subsubsection{Problems and gaps} In ubiquitous semantic Metaverse, the AR/VR applications have gained significant attention as potential next-generation technologies for contextual information provision and are being explored for their capabilities. Current AR/VR user interfaces for real-world browsing still have limitations, as they do not account for the variability in the meanings of objects depending on context~\cite{son2012contextual}. To address this issue, SIoT can be relied upon when addressing the interoperability and associated IoT-based contextual information provisioning, underlying design considerations and system architecture. For example, AR/VR devices can use an information exchange process that enables entities, not using a standard syntax or protocol, to accomplish tasks~\cite{wang2020developing,xie2020lite}. This is done by understanding each other's identities and communication objectives solely through universally shared knowledge, without the need for a dedicated intermediary or a global standard. 
Figure~\ref{fig:siot} illustrates the use of SIoT in addressing the challenges of IoT-based contextual information provisioning. The figure depicts the IoT ecosystem, which includes different devices and sensors that collect data and communicate with each other. SIoT provides a semantic layer that allows devices to understand the meaning of the data they collect and share. This semantic layer enables devices to communicate with each other and share data, even if they use different syntaxes or protocols.
The figure also shows how SIoT can address the variability in object meanings depending on context. 
By providing a shared understanding of the meaning of objects, SIoT can enable AR/VR devices to provide more accurate and relevant contextual information.

\subsubsection{Impacts} SIoT can be broken down into agent-specific semantic multiverses (SMs) for humans and machines~\cite{park2022enabling,zhang2022semantic}. Each agent's SM contains a semantic encoder and a generator, taking advantage of the AI and SIoT architecture. To enhance communication efficiency, the encoder acquires semantic representations (SRs) of multimodal data, while the generator learns to manipulate these representations for local scene rendering and interactions in SIoT. Given that these acquired SMs are biased towards local environments, their effectiveness depends on synchronizing diverse SMs in the background while communicating SRs in the foreground. This process involves aligning and reconciling the differences in knowledge, context, and understanding between various agents~\cite{xu2022edge}. SMs can enable SIoT devices to communicate and collaborate effectively, even when they have different data formats, protocols, or underlying technologies. This is achieved by relying on universally shared knowledge and understanding.

The latest generation of wireless technology, 5G and 6G, aims to support a wide range of SIoT by providing network capabilities for AR/VR users who share various content types, requirements, and semantics~\cite{dong2022edge,strinati20216g,chun2015semantic}. This has generated considerable interest in the ubiquitous semantic Metaverse. For example, the architectures based on federated edge intelligence for semantic-aware networking can be developed~\cite{shi2021semantic}. In this type of architecture, users of augmented and virtual reality can transfer the resource-intensive semantic processing tasks to edge servers. Multiple edge servers can cooperate in training a joint model for handling shared semantic knowledge, utilizing a federated learning-based structure.

\subsection{Use cases} SIoT can enable smart home systems to understand and interpret the connections between various AR/VR devices and their functionalities, allowing for more efficient automation and control. For example, an ontology-based approach can manage and coordinate energy consumption, security, and entertainment systems in a smart home~\cite{chen2019semantic,fensel2013sesame,chen2009semantic}. The SIoT can construct a model of a smart home's typical behavior using event logs and semantic data~\cite{son2013semanticradar}. For facilitating the functionality of AR/VR devices, the SIoT creates potential correlations based on semantic details such as applications, device categories, relationships, and installation sites, and then confirms these correlations using event logs. Correlations derived from the installed smart applications are employed to refine the mined correlations. A shadow execution engine can be developed to utilize these refined correlations to imitate the smart home's standard behaviors~\cite{fu2021hawatcher}. Inconsistencies between the actual states of devices and the simulated states are identified as anomalies during operation.

In a smart city context, SIoT can help manage and integrate various services and infrastructure, such as traffic control, waste management, public transportation, and energy distribution. Using semantic information, city systems can optimize resource usage, improve service efficiency, and facilitate better decision-making~\cite{an2019toward,balakrishna2020survey,zhang2022context}. The variety, origin, and structure of urban data are examined and standardized to enable data fusion across different modalities~\cite{rubi2021iot}. According to~\cite{balakrishna2018semantic,zhang2016semantic,d2015smart}, the smart city can be divided into sections, and the city data can be obtained for a specific section from a smart city's web services. It then becomes feasible to employ fusion technologies~\cite{lau2019survey} to learn latent features for each section, generating urban knowledge according to application requirements. Semantic technologies can model and annotate all data types, enhancing data value meaning while concealing data source and environment complexity by offering a standardized representation format. As an example, the SIoT is used in~\cite{solmaz2019toward} for managing crowd movement and crowd mobility analysis in Gold Coast, Australia, and Santander, Spain.

\begin{figure*}[h]
	\centering
	\includegraphics[width=0.9\textwidth]{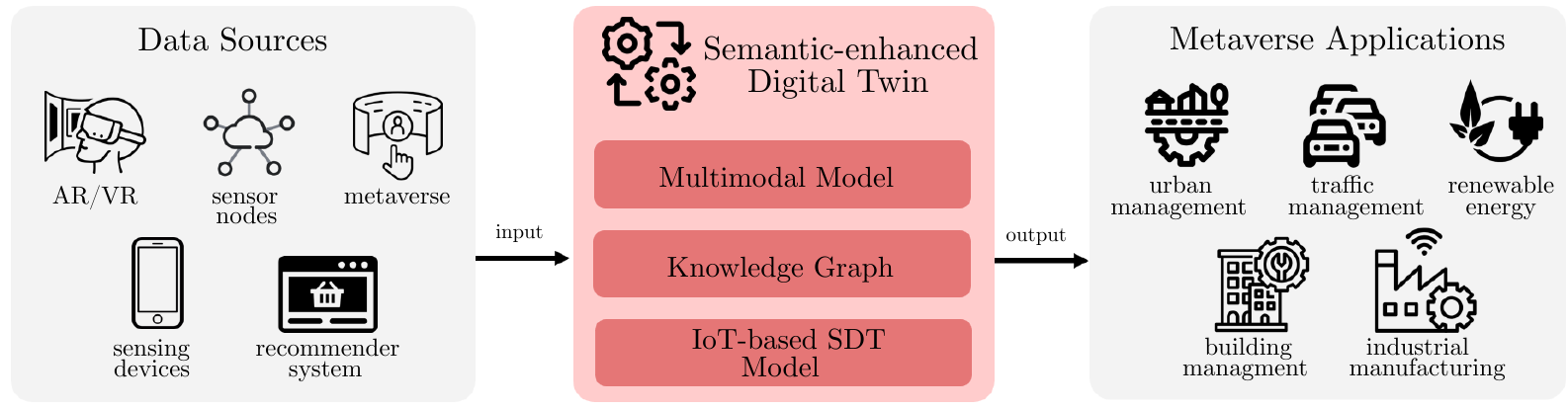} 
	\caption{A brief overview of the SDT that consists of a multimodal model, knowledge graph, and IoT-based SDT model.}
	\label{fig:sdt}
\end{figure*}

Moreover, SIoT can play a crucial role in healthcare by enhancing remote patient monitoring, telemedicine, and personalized health services. By utilizing semantic information, healthcare providers can better interpret and analyze patient data, track patients' conditions, and provide timely and accurate treatment recommendations~\cite{tiwari2020semantic,manonmani2020review,kovachev2014mobile}. The healthcare landscape comprises intricate interconnections among numerous technologies, processes, and individuals. Establishing robust communication between these components is essential for efficient healthcare management. SIoT can serve as knowledge representation instruments that employ abstractions to comprehensively define a specific subject, encompassing entities and their relationships~\cite{juneja2021iot,zgheib2020scalable,reda2019semantic}. The SIoT can also facilitate semantic annotation of data from diverse IoT devices, enabling annotations for the data. Resource description frameworks can be utilized for connecting entities using triples, making the information semantically meaningful~\cite{luschi2023semantic,balakrishna2020iot,jabbar2017semantic}. The annotated patient data also enhance semantic interoperability.

\section{Semantic-enhanced Digital Twin}
\label{sec_SDT}
In this section, we specifically investigate the representative techniques and use cases of SDT that creates virtual representations of physical objects, systems, or processes, enabling advanced simulations, analysis, and optimization within the ubiquitous semantic Metaverse.

\subsection{Representative techniques} 
\subsubsection{Problems and gaps} The digital twin, a digital representation of a physical system, is regarded as a practical and feasible method to tackle interactions between two AR/VR devices in the Metaverse due to its ability to offer perception and cognition capabilities. The digital twin has been extensively used for monitoring a variety of products, processes, and systems. However, there has been limited focus on utilizing DT within the context of ubiquitous semantic Metaverse. A brief overview of the SDT is depicted in Figure~\ref{fig:sdt}, which consists of a multimodal model, knowledge graph, and IoT-based SDT model. An SDT system based on a multimodal model can be used for monitoring real-time interactions and extracting latent semantic knowledge about these interactions~\cite{li2021semantic}. The entire system can be designed for lightweight operation, with the results displayed using AR/VR technology for real-time operation at an acceptable computational cost. The implementation of the SDT offers a feasible solution for monitoring both geometric and physical behaviors while achieving unified semantic mapping of device-environment interaction. This semantic mapping allows for the dynamic linking of entity attributes and relationship knowledge, providing a comprehensive understanding of the interactions~\cite{boje2020towards}.

\subsubsection{Impacts} Associating SDT system data collected through IoT sensors or AR/VR devices will be ineffective if the connections to the knowledge graph that represents the knowledge base of the SDT are not defined. Knowledge graphs can be used to support SDT that requires access to structured knowledge in the ubiquitous semantic Metaverse~\cite{gomez2019sedit}, which would distinctly define and interrelate the concepts of service time, data rate, and flow of production as a single unit. 
The knowledge graph can also be developed utilizing semantic web technologies~\cite{akroyd2021universal}. The system comprises ideas and examples delineated through ontologies, as well as computational agents that operate on these concepts and instances in order to update the evolving knowledge graph. 
In addition, knowledge graphs have emerged with the primary goal of consistently collecting data that describes dynamic real-world entities and their relationships in a unified model to gain new insights~\cite{sahlab2021knowledge}. Working inconspicuously within leading search engines and recommendation platforms such as Google, Amazon, and eBay, knowledge graphs are distinguished by their capacity to handle vastly diverse, constantly changing, and intricate information. 

An IoT-based SDT model~\cite{muralidharan2020designing} can be developed to represent an IoT device or application throughout its lifecycle virtually. The connection among the IoT devices can be enhanced by leveraging SDT with processing power and storage capacity at the edge, supported by docker images. To improve interoperability between the devices and applications, a W3C-based digital twin is utilized, independent of platform-specific IoT device requirements and networking protocols. The IoT-based SDT, in conjunction with microservices, can also seamlessly orchestrate multiple IoT devices and applications in a simplified manner~\cite{steindl2021semantic}.

\subsection{Use cases} By integrating semantic technologies with digital twins in SDT, a unified understanding of traffic data, public transportation systems, and road networks can be achieved. This information can be used to optimize traffic flow, predict congestion, improve public transportation routes, and facilitate urban planning~\cite{austin2020architecting}. SDT can also be used to monitor energy consumption in buildings and across the city. By modeling the interactions between energy sources, distribution networks, and consumers, energy usage can be better understood, allowing for the implementation of effective energy-saving measures and the integration of renewable energy sources~\cite{kwon2022semantic,hu2022method}. Moreover, SDT can be used to monitor air quality, noise levels, and other environmental factors in a smart city. SDT combines data from various sources and uses semantic technologies to understand the relationships between different factors, which identifies pollution sources, develops targeted mitigation strategies, and informs urban planning decisions.

SDT can model the interactions between the building's energy-consuming systems, such as heating, ventilation, air conditioning, lighting, and appliances, as well as external factors like weather conditions and occupancy patterns~\cite{donkers2021real,eneyew2022toward}. This enriched understanding facilitates the identification of inefficiencies, optimization of energy usage, and implementation of demand response strategies. SDT can enhance the interoperability between various building systems and components, facilitating seamless integration with building automation systems~\cite{jiang2023intelligent}, which allows efficient monitoring, control, and optimization of building performance. In particular, a specific project~\cite{de2020towards} is proposed to use SDT for the restoration of Notre-Dame de Paris with 3D digitization, where the assessment of spatial, temporal, and semantic intersections relies on the correlation of annotations, terminology, qualitative characteristics, and morphological aspects.

\begin{figure*}[h]
	\centering
	\includegraphics[width=0.9\textwidth]{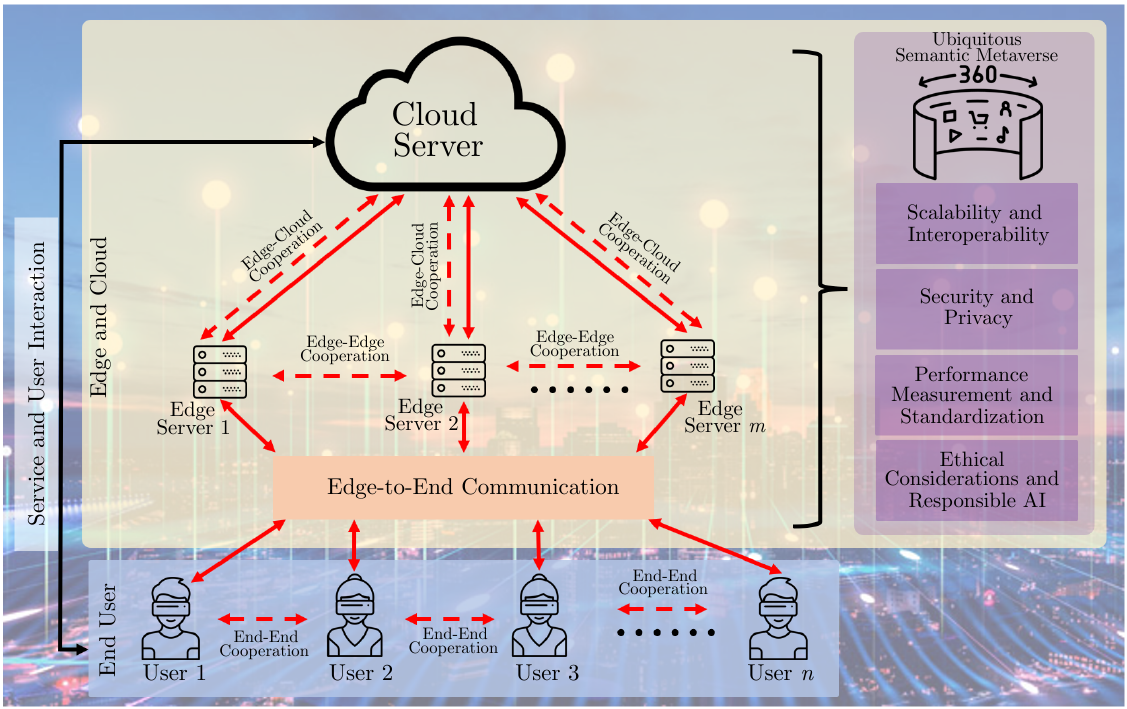} 
	\caption{An overview of future research directions for developing the ubiquitous semantic Metaverse with AI, STDR, SIoT and SDT.}
	\label{fig:futureDirection}
\end{figure*}

In industrial manufacturing, SDT can model the interactions between different production stages, equipment, and materials, enabling manufacturers to identify bottlenecks, inefficiencies, and potential improvements in the production process, which leads to increased productivity and reduced waste~\cite{zheng2020quality,perzylo2019opc}. SDT incorporates semantic information into factory digital twin, which enables manufacturers to better predict potential equipment failures, plan maintenance tasks proactively, and manage assets more effectively~\cite{redeker2021towards,li2021framework}. This approach can minimize downtime, reduce maintenance costs, and extend the lifespan of manufacturing equipment. 

\section{Research Opportunities For Future Ubiquitous Semantic Metaverse}
\label{sec_future}
We also investigate the research opportunities for constructing the future ubiquitous semantic Metaverse, including scalability and interoperability, privacy and security, performance measurement and standardization, ethical considerations and responsible AI, which are shown in Figure~\ref{fig:futureDirection}. Addressing those challenges is vital for creating a robust, secure, and ethically sound environment that offers engaging, immersive experiences for the users.

Based on the literature review in Sections~\ref{sec_AI} --~\ref{sec_SDT}, Table~\ref{tb_demands} summarizes the developing demands for satisfying the future research opportunities in terms of AI, STDR, SIoT, and SDT, where the ``red'' ${\color{red}\checkmark}$ indicates the most resources and supports are needed, the ``blue'' ${\color{blue}\checkmark}$ presents the fields requiring medium demands, and the ``black'' ${\color{black}\checkmark}$ is for the research with the least efforts. 

\begin{table}[htb]
\centering
\caption{Developing demands in terms of AI, STDR, SIoT, and SDT, where the ``red'' ${\color{red}\checkmark}$ indicates the most resources and supports are needed, the ``blue'' ${\color{blue}\checkmark}$ presents the fields requiring medium demands, and the ``black'' ${\color{black}\checkmark}$ is for the research with the least efforts.}
\label{tb_demands}
\begin{tabularx}{8cm}{l|XXXX}
\hline
& \bf{AI} & \bf{STDR} & \bf{SIoT} & \bf{SDT}  \\ \hline
Scalability and Interoperability & {\color{blue}\checkmark} & {\color{black}\checkmark} & {\color{red}\checkmark} & {\color{red}\checkmark} \\ 
& & & & \\
Security and Privacy & {\color{blue}\checkmark} & {\color{red}\checkmark} & {\color{red}\checkmark} & {\color{black}\checkmark} \\
& & & &  \\
Perform. Measure. and Standard. & {\color{red}\checkmark} & {\color{blue}\checkmark} & {\color{black}\checkmark} & {\color{red}\checkmark} \\
& & & &  \\
Ethical Consid. and Resp. AI & {\color{red}\checkmark} & {\color{red}\checkmark} & {\color{blue}\checkmark} & {\color{black}\checkmark} \\ \\
\hline
\end{tabularx}
\end{table}

\subsection{Scalability and Interoperability}
Enabling the ubiquitous semantic Metaverse with AI, STDR, SIoT and SDT requires the development of intelligent algorithms for allocating resources, such as computational power and bandwidth, based on real-time demand and user distribution. This has to maintain a high level of performance and user experience even as the ubiquitous semantic Metaverse continues to grow. Due to the fast querying and updating of objects, interactions, and relationships across the ubiquitous semantic Metaverse, investigating scalable data structures and indexing techniques is important to efficiently manage the vast amount of spatio-temporal information in the ubiquitous semantic Metaverse. Moreover, we need to design scalable networking and communication protocols to accommodate an increasing number of users, AR/VR devices, and data traffic within the ubiquitous semantic Metaverse. This will ensure that low-latency communication as well as the sharing of large datasets and complex interactions can be supported.

To enable interoperability between diverse ubiquitous semantic Metaverses, the research and development of AI, STDR, SIoT and SDT should focus on developing techniques for cross-domain semantic mapping and alignment. This includes methods for aligning and integrating heterogeneous ontologies, vocabularies, and data representations, as well as resolving inconsistencies and conflicts that may arise during the integration process. The interoperable communication protocols and APIs are also crucial for enabling seamless data exchange and AI interaction within the ubiquitous semantic Metaverse. In addition, interoperability in the ubiquitous semantic Metaverse requires the ability to process and reason over large-scale, heterogeneous, and dynamic data. Future research should focus on developing scalable and efficient algorithms for semantic data processing, AI, SIoT and SDT to support seamless integration and interaction according to the spatio-temporal AR/VR applications in the ubiquitous semantic Metaverse.

\subsection{Security and Privacy}
In the ubiquitous semantic Metaverse, data sharing and integration are essential for providing seamless experiences to users while maintaining their privacy. Envisioning privacy-preserving techniques for data sharing and integration in the Metaverse requires a comprehensive approach that addresses multiple layers of the system. This includes secure data storage and access, data anonymization, cryptographic techniques~\cite{li2019design}, federated learning~\cite{zheng2022exploring}, and decentralized identity management~\cite{zheng2021federated}. Data in the ubiquitous semantic Metaverse has to be stored using encryption techniques, ensuring that only authorized parties can access it. This can be achieved through secure multiparty computation, which enables parties to jointly compute a function over their inputs while keeping the inputs private. 

To safeguard user privacy, it is crucial to anonymize data prior to sharing or integration. Approaches, such as k-anonymity, l-diversity, and differential privacy, can be developed to ensure that the data cannot be traced back to individual users. Homomorphic encryption can enable computation on encrypted data, negating the requirement for decryption. This allows instantaneous data sharing and integration without disclosing the actual data to third parties in the ubiquitous semantic Metaverse. Moreover, the use of zero-knowledge proofs also allows for data verification without the exposure of its content.

Instead of sharing raw data, the ubiquitous semantic Metaverse can apply federated learning techniques to train machine learning models on decentralized data sources with AR/VR applications. This ensures that the semantic data remains private, and only the model updates are shared among participating parties, maintaining user privacy. To prevent identity theft and unauthorized data access, federated learning should employ decentralized identity management systems in the ubiquitous semantic Metaverse, where the blockchain technology~\cite{lin2023blockchain,khan2021blockchain,mollah2020blockchain,hassan2019blockchain} or other decentralized solutions are utilized to allow users to create self-sovereign identities, being verified without the need for central authorities. 

In the ubiquitous semantic Metaverse, security threats like eavesdropping, jamming, and virtuality-reality-synthesized issues~\cite{han2022paradefender,li2019energy} can pose significant risks to the privacy and integrity of user data. Attackers can intercept and manipulate semantic messages to deceive or disrupt the Metaverse experience~\cite{du2022rethinking}. AI algorithms can be further developed to identify and counter eavesdropping and jamming attacks. The ubiquitous semantic Metaverse should establish secure communication channels between users and AR/VR services, where end-to-end encryption is employed to ensure that only the intended sender and receiver can access the contents of the semantic messages. To protect against jamming attacks, spread spectrum techniques such as frequency hopping spread spectrum (FHSS) and direct sequence spread spectrum (DSSS) can be implemented in the ubiquitous semantic Metaverse, where the semantic signal is distributed over a wide range of frequencies or pseudo-random sequences are used to minimize the impact of interference on the communication channel.

\subsection{Performance Measurement and Standardization}
Performance measurement and standardization is vital for the heterogeneous AR/VR devices in the ubiquitous semantic Metaverse, which ensures seamless and high-quality user experiences across different hardware platforms. A comprehensive approach to standardizing spatio-temporal performance should address device capabilities, interoperability, and application requirements. Establishing standardized profiles for different device capabilities can optimize the AR/VR applications for various hardware specifications that include aspects like rendering capabilities, tracking precision, display resolution, and field of view. To facilitate seamless semantic communication and data exchange between heterogeneous AR/VR devices, interoperability standards should be established. These standards can define protocols, data formats, and APIs that enable efficient integration and interaction between different devices and platforms in the ubiquitous semantic Metaverse. 

Encouraging the development of cross-platform applications and tools provides consistent user experiences across multiple ubiquitous semantic Metaverses. Those applications and tools are required to adapt to various AR/VR device capabilities, thus promoting the use of standardized APIs, libraries, and development frameworks is important. We also need to design QoE metrics specific to the AR/VR devices, which evaluate the spatio-temporal performance of applications and services on different hardware platforms. These metrics may include factors such as latency, frame rate, tracking accuracy, and user comfort~\cite{qin2021semantic}. Moreover, implementing adaptive content delivery techniques can enable the ubiquitous semantic Metaverse to dynamically adjust the quality of AR/VR experiences based on the capabilities of the user's device, such as adjusting the resolution, level of detail, or rendering quality depending on the available processing power and bandwidth. 

\subsection{Ethical Considerations and Responsible AI}
The privacy of user data has to be guaranteed while maintaining transparency regarding data collection, storage, and usage is vital. Users should have control over their personal information and understand how it is used within the ubiquitous semantic Metaverse, particularly when it comes to their interactions within spatio-temporal environments. Users should have the option to maintain anonymity or use pseudonyms to protect their identities. At the same time, mechanisms should be in place to prevent malicious activities or harassment by anonymous users, especially as they navigate through various spaces and times in the ubiquitous semantic Metaverse. Moreover, the ubiquitous semantic Metaverse needs to be designed to accommodate users with diverse backgrounds, abilities, and preferences, ensuring equitable access to digital experiences and opportunities. This includes considering accessibility guidelines for users with disabilities and promoting cultural diversity and representation within spatio-temporal environments. 

As the ubiquitous semantic Metaverse is a spatio-temporal environment, ethical considerations should also address the dynamic nature of the virtual world, such as preserving the history and context of spaces and objects, ensuring that changes in the virtual environment respect cultural heritage, and managing the implications of time-travel or alternate reality experiences. The development and maintenance of the ubiquitous semantic Metaverse can consider the potential environmental impact of the technology, including energy consumption, electronic waste, and resource utilization. Developers should aim to minimize the ecological footprint and promote sustainable practices within the ubiquitous semantic Metaverse. AI and machine learning systems have to be designed and trained to minimize biases that may perpetuate discrimination or inequality. The techniques can be developed to mitigate biases in algorithms and datasets and regularly audit their systems to ensure fairness, particularly when analyzing and predicting spatio-temporal patterns.

Responsible AI can be applied to the ubiquitous semantic Metaverse, where the AI systems prioritize ethical considerations, user well-being, and long-term societal impact while taking into account the unique spatio-temporal aspects of the environment. Responsible AI in the ubiquitous semantic Metaverse needs to minimize biases and promote equitable treatment of all users, regardless of their background or characteristics, which may impact spatio-temporal analysis, recommendation systems, and user interactions. Moreover, responsible AI in the ubiquitous semantic Metaverse should prioritize transparency and explainability, allowing users to understand how AI systems make decisions and process spatio-temporal information. This can provide clear documentation, visualizations, and user interfaces that explain AI system behavior and outcomes. Consider security and privacy in the ubiquitous semantic Metaverse. Responsible AI should balance the users' privacy and adhere to data protection regulations for privacy-preserving techniques, such as federated learning, differential privacy, and secure multiparty computation, especially when processing sensitive spatio-temporal data.

\section{Conclusion}
\label{sec_conclusions}
This paper studies the intelligence and spatio-temporal characteristics of four fundamental components, i.e., AI, STDR, SIoT, and SDT, in the ubiquitous semantic Metaverse. By utilizing the four essential technologies, immersive cyber-virtual applications and anywhere-and-anytime services can be connected with the ubiquitous semantic Metaverse, striving to create intelligent, customized, real-time, and context-aware interactions for the users. We comprehensively survey the representative techniques for each of the components, such as NLP, semantic communications, graph-based models, agent-specific SM, IoT-based SDT model, etc., with their typical use cases. Moreover, we highlight the prospects and forthcoming research avenues for building an ubiquitous semantic Metaverse, encompassing aspects such as scalability and interoperability, privacy and security, performance assessment and standardization, along with ethical concerns and responsible AI implementation. Comprehending these opportunities and research directions is crucial in fostering the advancement of versatile cross-platform applications and tools that can cater to a range of AR/VR device capabilities, ensuring uniform user experiences throughout numerous future ubiquitous semantic Metaverse.

\section*{Acknowledgements}
This work was supported by the CISTER Research Unit (UIDP/UIDB/04234/2020), project ADANET (PTDC/EEICOM/3362/2021) and project IBEX (PTDC/CCI-COM/4280/2021), financed by National Funds through FCT/MCTES (Portuguese Foundation for Science and Technology).

\bibliographystyle{IEEEtran}
\bibliography{bibSCM}

\end{document}